\documentclass[a4paper,11pt]{article}
\usepackage{amsmath,amsfonts,amssymb,times,mathptmx,graphicx,amsthm,lineno}
\usepackage{placeins}
\usepackage[T1]{fontenc}
\usepackage[latin1]{inputenc}
\title{\bf Probability Distributions of Positioning Errors for Some Forms of
Center-of-Gravity Algorithms. Part II
 }
\author{Gregorio Landi$^a$\thanks{Corresponding
author. Gregorio.Landi@fi.infn.it}~,   Giovanni E. Landi$^b$\\
\\
\llap{$^a$} Dipartimento di Fisica e Astronomia,
Universita' di Firenze and INFN\\
Largo E. Fermi 2 (Arcetri) 50125, Firenze, Italy\\
\\
\llap{$^b$} ArchonVR S.a.g.l.,\\
Via Cisieri 3,
6900 Lugano, Switzerland.\\ \\
{ November 23, 2020}}
\date{ }
\begin{document}
\maketitle 
\begin{abstract}

The center of gravity is one of the most frequently used algorithm
for position reconstruction with different analytical forms for
the noise optimization.
The error distributions of the different forms
are  essential instruments to improve the track fitting in particle
physics. Their Cauchy-(Agnesi) tails have a beneficial
effects to attenuate the outliers disturbance in the
maximum likelihood search.
The probability distributions are calculated for
some combinations of random variables, impossible to
find in literature, but relevant for track fitting:
 $x_{g3}=\theta(x_2-x_1)[ (x_1-x_3)/(x_1+x_2+x_3)] +
 \theta(x_1-x_2)[(x_1+2x_4)/(x_1+x_2+x_4)]$
and $x_{g4}=\theta(x_4-x_5)[(2x_4+x_1-x_3)/(x_1+x_2+x_3+x_4)]+
\theta(x_5-x_4)[(x_1-x_3-2x_5)/(x_1+x_2+x_3+x_5)]$
and $x_{g5}=(2x_4+x_1-x_3-2x_5)/(x_1+x_2+x_3+x_4+x_5)$.
The probability density functions
of $x_{g3}$, $x_{g4}$ and $x_{g5}$ have complex structures
with regions of reduced probability. These regions must
be handled with care to avoid false maximums
in the likelihood function.
General integral equations and detailed analytical
expressions are calculated assuming the set $\{x_i\}$ as
independent random variables with Gaussian probability
distributions.
\end{abstract}

\newpage


\tableofcontents

\pagenumbering{arabic} \oddsidemargin 0cm  \evensidemargin 0cm


\section{Introduction }

This work continues the calculations~\cite{landi10,landi11} of
the error probability density functions (PDFs) for the
Center-of-Gravity (COG)
\footnote{Or "$\kappa\acute{\varepsilon}\nu\tau\rho\alpha\ \ \beta\alpha\rho\widetilde{\omega}\nu$"
in the Archimedes treatise~\cite{archimede}  or
"$centrum$ $gravitatis$" in G. Galilei~\cite{galileo}.}
in its discretized forms.
These COG expressions are geometrical tools
of large use in particle physics as easy and
efficient positioning algorithms.
Their generic definitions are synthesized as: $X_{g}=\sum_jx_j\tau_j/\sum_jx_j$,
where $x_j$ are the signals of a cluster inserted in $X_{g}$
and $\tau_j$ their positions.
To filter the noise, the COG algorithm has different forms based on the number of
discrete data used in $X_g$. In our case, we consider hits on  particle detectors,
each hit distributes a signal in a restricted number (four of five at the most)
of nearby sensor substructures (strips in silicon micro-strip detectors). The number of strips
used in the COG algorithm have important effects on the corresponding positioning
errors. For example, in the case of orthogonal incidence of refs.~\cite{landi05,landi06},
the two strip COG suffices to produce excellent results. Instead the three strip COG
has a consistent lower performance due to the noise of the third strip. However,
at larger angles the signals of the third strips are relevant. The addition of a
third (or forth) strip has a drastic effect in the expressions
of the PDFs, as easily observed in the corresponding COG
histograms~\cite{landi05,landi01}. Discontinuities
or gaps in the PDFs are present. The COG algorithms with an even number of strips have
gaps around zero, algorithms with an odd number of strips have gaps around the strip borders.
The gap dimensions are tuned by the sizes of the
signal distributions~\cite{landi01,landi03}.
This fact obliges to a separate study of each selected strip number.
The beneficial effects of these PDFs in track reconstructions are
reported in refs.~\cite{landi05,landi06,landi07},
where extensive use of the simplest of these PDfs was done in
the maximum likelihood search or the simpler schematic model
introduced to initialize the maximum likelihood search.
The results of ref.~\cite{landi06} were illustrated in
ref.~\cite{landi07} with a very simple Gaussian model where
the complex structure of the hit variances was reduced to
only two types of variance (minimal heteroscedasticity).
Further extensions of this model are discussed in
ref.~\cite{rudi}. To prove beyond any possible doubt
the goodness of the results of the schematic model of
ref.~\cite{landi07}, general demonstrations are
reported in refs.~\cite{landi08,landi09}, these
demonstrations will save also the results of the maximum
likelihood from similar doubts.
The first forms of the PDFs of the COG algorithm
are explored in refs.~\cite{landi10,landi11} with
two different methods of calculation.
Here we will follow the shorter method of ref.~\cite{landi10}.
The longer method of ref.~\cite{landi11}, applied to this type
of problems, will be reported elsewhere
to complete the PDFs with their cumulative distribution
functions. The use of MATHEMATICA~\cite{MATHEMATICA}
is essential for the construction of the analytical
expressions of the PDFs and the numerical integrations
to test the approximate analytical results.
The simulations are produced with MATLAB~\cite{matlab}.

\section{The three strip COG (COG$_3 $) PDF and the border gaps }

As anticipated at the end of ref.~\cite{landi10}, the complete form of the
the PDF for the COG$_3$ algorithm must account the noise
effect that promotes an adjacent strip to
become the seed strip (the strip with the maximum signal of a cluster). In
general, we assume that the origin of our reference system is in the
center of the strip crossed by the particle. This strip
has an high probability to be the seed strip, but the noise fluctuation
can modify this condition at the strip border. For an even number of strips
this noise effect is irrelevant for the continuity of the PDF at the
strip borders. For odd numbers of strips, gaps are produced in
the COG histograms at the borders (ref.~\cite{landi01,landi03}).
These gaps must be carefully considered in the maximum likelihood
search for their relevant modifications of the PDFs.

\subsection{The complete form of the COG$_3$ PDF at the right border}

To save a backward consistency with the conventions of
refs.~\cite{landi10,landi11}, the strip numbering
becomes very peculiar.
The cluster of five strips, used in the following,
are indicated with the numbers: $ 5,3,2,1,4$. The origin of the
reference system is always in the center of the strip $\#2$.
The strip width is the unity of length. The most probable triplet of strips
is supposed to be $3,2,1$ with the seed on strip $\#2$.
But, near to the right borders,  the noise can promote the strip $\#1$
(or the strip $\#3$ in the left borders) to be the seed, and the triplet
of the COG$_3$ becomes $2,1,4$.
\begin{equation}\label{eq:equation_1}
    x_{g_3}=\Big(\frac{x_1-x_3}{x_1+x_2+x_3}\Big)\theta(x_2-x_1)+
    \Big(\frac{x_4-x_2}{x_1+x_2+x_4}+1\Big)\theta(x_1-x_2) \,.
\end{equation}
\noindent
The COG$_3$ for the triplet $2,1,4$ in the reference system on the
strip $\#2$ is $(x_1+2 x_4)/(x_1+x_2+x_4)$ but the form of
equation~\ref{eq:equation_1} will be useful in the following.
With the method of refs.~\cite{landi10,landi06} this PDF becomes
\footnote{If the variables $\{x_i\}$ are correlated, the modification of
this equation implies a single PDF for all the variables}:
\begin{equation}\label{eq:equation_2}
\begin{aligned}
   &P_{xg_3}(x)= \int_{-\infty}^{+\infty}\,\mathrm{d} x_1\,
   \mathrm{d} x_2\, \mathrm{d} x_3\, \mathrm{d} x_4 P_1(x_1) P_2(x_2) P_3(x_3) P_4(x_4)\\
   &\Big\{\delta\big(x-\frac{x_1-x_3}{x_1+x_2+x_3}\big)\theta(x_2-x_1)+
   \delta\big(x-(\frac{x_4-x_2}{x_1+x_2+x_4}+1)\big)\theta(x_1-x_2)\Big\}\,.
\end{aligned}
\end{equation}
\noindent
The normalization of $P_{xg_3}(x)$ is immediately verified with a
direct integration on $x$ of the Dirac $\delta$-functions, the
normalization of the remained PDFs $P_j(x_j)$ assures the result.
The integrals of
equation~\ref{eq:equation_2}, with the first $\delta$-function,
differ from those of ref.~\cite{landi10} by a function
$\theta(x_2-x_1)$, but they can be processed
in a similar way. The substitution of variables
$\xi=(x_1-x_3)$,  $z'=(x_1+x_2+x_3)$ and $\beta'=x_2$
simplifies the integration of the $\delta$-function.
The Jacobian-determinant of this substitution
is $1/2$. Integrating in $\xi$ the $\delta$-function, the
remaining double integral has the following form:

\begin{equation}\label{eq:equation_3}
    P_{xg_3}^a(x)=\frac{1}{2}\int_{-\infty}^{+\infty}\,\mathrm{d}z'\,|z'|\,
    \int_{z'(1+x)/3}^{+\infty}\mathrm{d}\beta'  P_1(\frac{z'(1+x)-\beta'}{2})
    P_2(\beta')\,P_3(\frac{z'(1-x)-\beta'}{2})\,.
\end{equation}
\noindent
The $\xi$ integration eliminates the factor $1/x^2$, always present in
ref.~\cite{landi10} to remember the Cauchy-like tails. Evidently, the PDF
continues to have the Cauchy-like tails, but now the apparent
singularity $1/x^2$ for small $x$ is automatically suppressed.
With another substitution of variables, $z'=3z$ and $\beta'=2\beta+z(1+x)$, the
$\beta$ integration has zero as lower limit and
equation~\ref{eq:equation_3} can be recast in the form:

\begin{equation}\label{eq:equation_4}
    P_{xg_3}^a(x)=9\int_{0}^{+\infty}\mathrm{d}\beta
    \int_{-\infty}^{+\infty}\,\mathrm{d}z\,|z|\,P_1(z(1+x)-\beta)
    P_2(2\beta+z(1+x))\,P_3(z(1-2x)-\beta)\,.
\end{equation}
\noindent
The integrals of the second $\delta$-function of
equation~\ref{eq:equation_3} can be reduced to the first
ones with a set of variable substitutions. If $x_2\rightarrow x_1$,
$x_1\rightarrow x_2$, $x_3\rightarrow x_4$ and $x\rightarrow 1-x$,
the integrals of equation~\ref{eq:equation_4} become those of the second part
of equation~\ref{eq:equation_3}. Thus, this part of the PDF is recovered
from the previous integrals. Let us carefully illustrate these
substitutions that will be used often in the following:
\begin{equation*}
\begin{aligned}
      & x_2\rightarrow x_1 \ \ \ \ \ x_1\rightarrow x_2 \ \ \ \ x_3\rightarrow x_4 \\
     &\delta\big(x-\frac{x_1-x_3}{x_1+x_2+x_3}\big)\theta(x_2-x_1)\rightarrow
     \delta\big(x-\frac{x_2-x_4}{x_1+x_2+x_4}\big)\theta(x_1-x_2)\rightarrow\\
     &\delta\big(-x-\frac{x_4-x_2}{x_1+x_2+x_4}\big)\theta(x_1-x_2)\rightarrow
     \delta\big(1-x-\frac{x_4-x_2}{x_1+x_2+x_4}-1\big)\theta(x_1-x_2)
\end{aligned}
\end{equation*}
\noindent
In the last term of the second line,  all the signs
are changed for the
symmetry of the $\delta$-function obtaining the first term
of the third line. Adding and subtracting one and substituting
$x\rightarrow 1-x$, the last $\delta$-function becomes
the corresponding one of equation~\ref{eq:equation_2}.
The equations for $P_{xg_3}^a(x)$ are exact, but
the integrals with Gaussian PDFs requires numerical integrations,
too slow for a maximum likelihood search.
Approximate analytical expressions are fundamental,
even if their final forms depend from the used approximations.
However, the main parts remain very similar, the
differences are in the small terms that add negligible modifications
to the tails of the PDFs (always Cauchy-like). The
essential effects of these tails are to softly suppress
the outlier hits, as illustrated in ref.~\cite{landi05}.

\subsection{Partial expression of the COG$_3$ PDF with Gaussian noise}

As in ref.~\cite{landi10}, an additive Gaussian noise
will be supposed to perturb the signal $a_i$ collected by the strip $i$
with a standard deviation $\sigma_i$:

\begin{equation}\label{eq:equation_5}
    P_i(z)=\frac{1}{\sqrt{2\pi}\,\sigma_i}\exp\Big[-\frac{(z-a_i)^2}
    {2\sigma_i^2}\Big]\ \ \ \ i=1,2,\cdots
\end{equation}
\noindent
With this type of noise, the double integration of
equation~\ref{eq:equation_4} has no analytical
form, and approximations must be constructed. The
method of approximation of ref.~\cite{landi10} works
very well even in this case. Given that the probabilities
$P_i$ are Gaussian, so are their products. Hence,
as first step, we integrate by parts in $z$, completing
the $z$-factor adding and subtracting a term to construct
the derivative  of the products of Gaussian $P_i$.
This additional term  is constant in $z$, giving Gaussian
integrals from zero to $+\infty$ and from $-\infty$ to
zero. The sum of these two integrals introduces an
Erf-function that excludes a closed form for the
last integration from zero to $\infty$ in $\beta$.
We neglect this Erf-function and integrate the rest
in $\beta$. The Erf-function assures the positivity
of the argument of the second integration, but in
practical applications that argument is always positive,
or an absolute value can be easily inserted, as
done in ref.~\cite{landi10}.
The $\beta$-integration adds another Erf-function,
this Erf-function is essential to originate the gap in the PDF.
Introducing the auxiliary variables:

\[X_3=(a_1-a_3)/(a_1+a_2+a_3) \ \ \ \ \ \ E_3=a_1+a_2+a_3\,,\]
\noindent
the first part of the PDF, $P_{xg_{3}}^a(x)$, becomes:

\begin{equation}\label{eq:equation_6}
\begin{aligned}
    P_{xg_{3}}^a(x)=&\left\{\exp\Big[-(X_3-x)^2\frac{E_3^2}
    {2[(1-x)^2\sigma_1^2+x^2\sigma_2^2+(1+x)^2\sigma_3^2]}\Big]\right\}\\
    &\frac{\,\,\Big|E_3\,[(1-X_3)(1-x)\sigma_1^2+X_3x\sigma_2^2+(1+X_3)(1+x)
    \sigma_3^2]\Big|\,\,}
    {\sqrt{2\pi}\Big[(1-x)^2\sigma_1^2+x^2\sigma_2^2+(1+x)^2\sigma_3^2\Big]^{3/2}}
    \,\frac{\big[1+\mathrm{Erf}(H)\big]}{2},\\
\end{aligned}
\end{equation}
\noindent
Where H is define as:
\begin{equation*}
      H=\frac{a_1(1-2x)x\sigma_2^2-a_3(1+x)x\sigma_2^2+a_3(x^2-1)
      \sigma_1^2+a_2(1-3x+2x^2)\sigma_1^2-(a_1-a_2)(1+x)^2\sigma_3^2}
       {\sqrt{2}\sqrt{(1+x)^2\sigma_3^2(\sigma_1^2+\sigma_2^2)+
       \sigma_1^2\sigma_2^2(1-2x)^2}
       \sqrt{(1-x)^2\sigma_1^2+x^2\sigma_2^2+(1+x)^2\sigma_3^2}}\,.
\end{equation*}
\noindent
The equation~\ref{eq:equation_6} is very similar to the PDF
of the simplified PDF for the COG$_3$ of ref.~\cite{landi10}, with the
only evident modification of the $[1+\mathrm{Erf}(H)]/2$. This
factor introduces the gap in the PDF. It operates a
smooth transition with the other part of the PDF, dominated by the strip
$\#1$ as seed strip, and it reduces the amplitude of each part.
Writing $P_{xg_{3}}^a(x)$ with all its parameters, the complete PDF is:
\begin{equation}\label{eq:equation_7}
    P_{xg_{3}}(x,a_1,a_2,a_3,a_4,\sigma_1,\sigma_2,\sigma_3,\sigma_4)=
    P_{xg_{3}}^a(x,a_1,a_2,a_3,\sigma_1,\sigma_2,\sigma_3)+
    P_{xg_{3}}^a(1-x,a_2,a_1,a_4,\sigma_2,\sigma_1,\sigma_4)
\end{equation}
\noindent

\begin{figure} [h!]
\begin{center}
\includegraphics[scale=0.65]{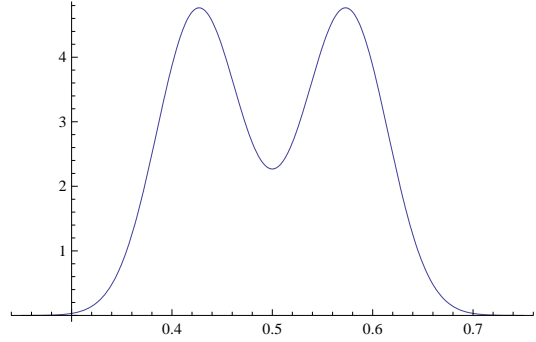}
\caption{\em  $P_{xg3}(x)$ around $x=1/2$ (blue line) overlapping
the numerical integration (red line).}
\label{fig:figure_1}
\end{center}
\end{figure}

The other (small) term is expressed by:

\begin{equation}
\begin{aligned}
    P_{xg_{3}}^s(x)=&\exp\Big[\frac{(a_1-a_3)^2(1+x)^2\sigma_3^2+
    (a_3-a_1+(2a_1+a_3)x)^2\sigma_2^2+(a_2-a_3-(2a_2-a_3)x)^2\sigma_1^2}
    {2((1-2x)^2\sigma_1^2\sigma_2^2+(1+x)^2(\sigma_1^2+\sigma_2^2)\sigma_3^2)}\Big]\\
    &\frac{(1+x)\sigma_2^2\sigma_3^2-2(1+x)\sigma_1^2\sigma_3^2+(1-2x)\sigma_1^2\sigma_2^2}
    {2\pi\sqrt{(1+x)^2(\sigma_2^2+\sigma_1^2)\sigma_3^2+(1-2x)^2\sigma_1^2\sigma_2^2}
    \big(\sigma_1^2(1-x)^2+x^2\sigma_2^2+(1+x)^2\sigma_3^2\big)}\\
    &\mathrm{Erf}\Big[\frac{a_1(1+x)\sigma_2^2\sigma_3^2-2a_2(1+x)
    \sigma_1^2\sigma_3^2+a_3(1-2x)\sigma_1^2\sigma_2^2}
    {\sigma_1\sigma_2\sigma_3\sqrt{2}\sqrt{(1+x)^2(\sigma_2^2+\sigma_1^2)
    \sigma_3^2+(1-2x)^2\sigma_1^2\sigma_2^2}
    }\Big]\,.
\end{aligned}
\end{equation}
\noindent
It contains the Erf-function neglected in $P_{xg_{3}}^a(x)$ and it is not far to the
form  $A\,\mathrm{Erf}(A)$ even if now the term in the Erf-function differs by
the positive constants $\{a_j\}$. Near the strip borders, two of the constants $a_i$
are very similar and this term is positive (even if negligible).

\noindent
The last term, indicated as Cauchy term, is:

\begin{equation}
\begin{aligned}
    P_{xg_3}^C(x)=&\exp\big[-\frac{E_3^2\,[\sigma_1^2(X_3-1)^2+
    \sigma_2^2(X_3)^2+\sigma_3^2(X_3+1)^2]}
    {2(\sigma_1^2\sigma_2^2+4\sigma_1^2\sigma_3^2+\sigma_3^2\sigma_2^2)}\big]\\
    &\Big\{1+\mathrm{Erf}\big[\frac{-a_1\sigma_2^2\sigma_3^2+
    2a_2\sigma_1^2\sigma_3^2-a_3\sigma_1^2\sigma_2^2}
    {\sqrt{2\sigma_1^2\sigma_2^2\sigma_3^2(\sigma_2^2\sigma_3^2+4\sigma_1^2\sigma_3^2+
    \sigma_1^2\sigma_2^2)}}]\Big\}\\
    &\Big\{\frac{{9\sigma_1^2\sigma_2^2\sigma_3^2}}
    {2\pi\sqrt{\sigma_2^2\sigma_3^2+4\sigma_1^2\sigma_3^2+\sigma_1^2\sigma_2^2}
    [(1-2x)^2\sigma_1^2\sigma_2^2+(1+x)^2(\sigma_1^2+\sigma_2^2)\sigma_3^2]}\Big\} \,.\\
\end{aligned}
\end{equation}
\noindent
This term is
different from zero also when all the $a_j=0$. In this case the
remaining term is the last line of the previous equation.

\subsection{Examples and simulations}

\begin{figure} [h!]
\begin{center}
\includegraphics[scale=0.8]{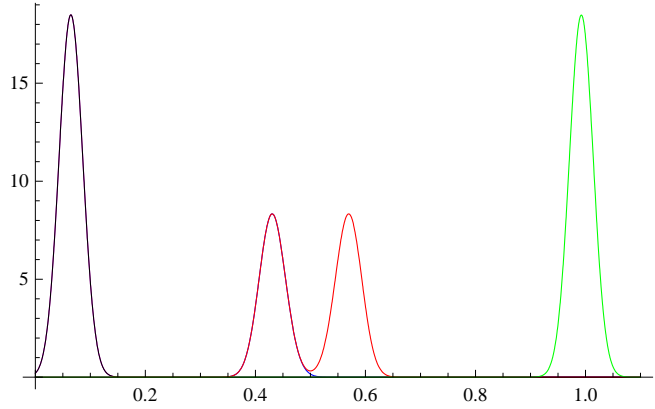}
\caption{\em  Samples of  $P_{xg3}(x)$ and $P_{xg3}^a(x)$
for various values of the parameters $a_i$.
The black line shows a coincides $P_{xg3}^a(x)$ for $x\approx 0.5$
($a_4$ has no effect here). The blue line is $P_{xg3}^a(x)$
and it describes the main maximum. The red line is
the complete result ($P_{xg3}(x)$). The green line has the
$a_i$-parameters for $x\approx 1$ (plots with MATHEMATICA).
}\label{fig:figure_2}
\end{center}
\end{figure}

Figure~\ref{fig:figure_2} illustrates the various
forms of $P_{xg3}(x)$ and its splitting in two maximums around
$x\approx 1/2$. For $x\approx 0$ or $x\approx 1$ the second
maximum disappears.

\begin{figure} [h!]
\begin{center}
\includegraphics[scale=0.4]{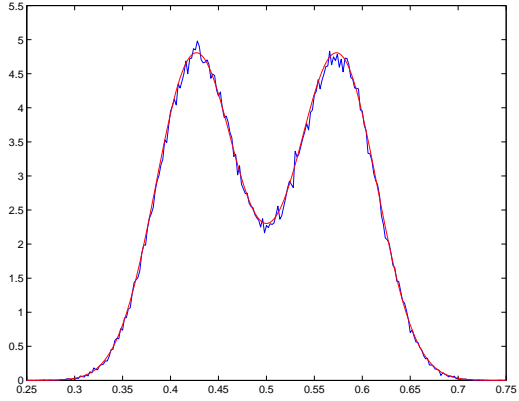}
\caption{\em  Comparison of  $P_{xg3}(x)$ (red line) with the simulations,
the noiseless energies are  $\{a_1=66.5,a_2=66.5,a_3=5.17,a_4=5.17\}$ and
$\sigma_i=3.9$ (ADC counts)
}\label{fig:figure_3}
\end{center}
\end{figure}

\noindent
Some simulations are constructed to test the results of
the calculation. The parameters $a_i$ are from ref.~\cite{landi03}.
Gaussian random noises are added to the $\{a_i\}$ to obtain the random
values $\{x_i\}$. The $\{x_i\}$ are inserted in equation~\ref{eq:equation_1}
and the empirical PDF is compared with the analytical results.
The simulations of figure~\ref{fig:figure_3} show a nice agreement.

\begin{figure} [h!]
\begin{center}
\includegraphics[scale=0.4]{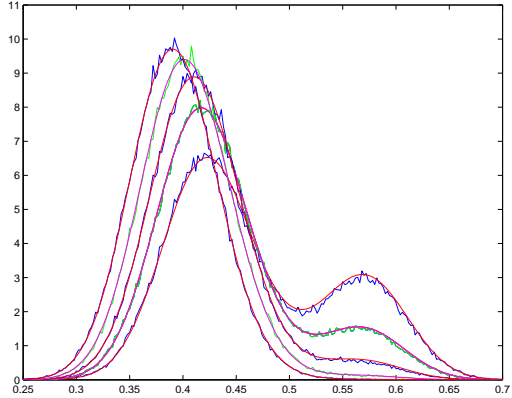}
\caption{\em  The PDF $P_{xg3}(x)$ and the simulations with
200000 events for impact points $\{0.48,\,0.46,\,0.44,\,0.42,\,0.40\}$.
The second maximum rapidly disappears.
}\label{fig:figure_4}
\end{center}
\end{figure}

The illustration of the soft switch-off of the second maximum is more difficult, because a
detailed model of the modifications of the parameter $a_i$ are required.
The trends of the $a_i$ of ref.~\cite{landi05} could be used in figure~\ref{fig:figure_4}.
Symmetrically the maximum of $P_{xg3}^a(x)$ rapidly disappears for impact points
greater than $0.5$, in the strip~$\# 1$
\newpage

\subsection{The complete form of the COG$_3$ PDF at the left border}

The effects of the noise described for the right border of the strip $\#2$ are
present even to the left border. Also here, the noise can promote the strip $\#3$
to become the seed strip. In this case the triplet of the algorithm become $5,3,2$,
the strips $\#5$ and the strip $\#2$ are the two lateral strips to the strip $\#3$,
This part of the COG$_3$ PDF becomes:

\begin{equation}\label{eq:equation_10}
    x_{g_3}^L=\Big(\frac{x_1-x_3}{x_1+x_2+x_3}\Big)\theta(x_2-x_3)+
    \Big(\frac{x_2-x_5}{x_2+x_3+x_5}-1\Big)\theta(x_3-x_2) \,.
\end{equation}
\noindent
As above, the second part of $x_{g_3}^L$ has its standard form in the reference
centered on the strip $\#3$, but the subtraction of one reports
its form to that centered on the strip $\#2$ i.e. $(-x_3-2 x_5)/(x_2+x_3+x_6)$.
With the usual constraints of ref.~\cite{landi10}, this PDF becomes:
\begin{equation}\label{eq:equation_11}
\begin{aligned}
   &P_{xg_3}^L(x)= \int_{-\infty}^{+\infty}\,\mathrm{d} x_1\,
   \mathrm{d} x_2\, \mathrm{d} x_3\, \mathrm{d} x_5 P_1(x_1) P_2(x_2) P_3(x_3) P_5(x_5)\\
   &\Big\{\delta\big(x-\frac{x_1-x_3}{x_1+x_2+x_3}\big)\theta(x_2-x_3)+
   \delta\big(x-(\frac{x_2-x_5}{x_2+x_3+x_5}-1)\big)\theta(x_3-x_2)\Big\}\,.
\end{aligned}
\end{equation}
\noindent
The appropriate substitutions allow the use of equation~\ref{eq:equation_3}
in these two new branches of the PDF. The first set of
substitutions are: $x_1\rightarrow x_3$, $x_3\rightarrow x_1$ and $x\rightarrow -x$.
These are essential to change the $\theta(x_2-x_1)$
of equation~\ref{eq:equation_3} in the $\theta(x_2-x_3)$ of equation~\ref{eq:equation_11}.
The second part of equation~\ref{eq:equation_11}
is obtained from  equation~\ref{eq:equation_3} with the substitutions:
$x_1\rightarrow x_2$, $x_3\rightarrow x_5$, $x_2\rightarrow x_3$ and
$x\rightarrow x-1$. Equation~\ref{eq:equation_7} shows the practical
use of these substitutions.

\subsection{The use in the maximum likelihood search}

The various sections of $P_{xg_3}^{R|L}(x)$ are relevant for the maximum
likelihood search, even if they are invisible in the COG$_3$ histograms. Typically,
the experimental COG$_3$ histograms have drops of data density
near to $\pm 1/2$ or absence of data (gaps).
The sectors of probabilities above $1/2$ or below $-1/2$
have few or no data. Instead, our $P_{xg_3}^{R|L}(x)$ has a different
aspect. This is due to the selection of the impact point as the
event producing the noiseless signals $\{a_j\}$ modified by the noise.
In the data, the strip of the hit is always supposed to be the
seed of the cluster. Thus, the forms of the COG$_3$ algorithm is always
$(x_1-x_3)/(x_1+x_2+x_3)$, where $3,2,1$ are the ordering of the strip signals
around the seed strip $\# 2$. If near to the right-border of the strip,
the noise promotes the nearby strip to becomes the seed strip, the COG$_3$
becomes negative and greater than $-1/2$, instead of being greater than $1/2$
as calculated in $P_{xg_3}^{R}(x)$. These apparent inconsistencies are
irrelevant for the maximum likelihood search. The PDF, explored in the
search, has always a fixed $x$-value, given by the COG$_3$ of the hit,
and it is tested  through the functions $\{a_j(\varepsilon)\}$ to
find the most probable impact point. Hence, the search of the
maximum of the PDF products must access the side of
$P_{xg_3}^{L}(x)$ in the nearby strip with a continuous transition.
The complex structure of the $P_{xg_3}^{R|L}(x)$ assures this
continuity.

\section{The error PDF of the four Strip COG (COG$_4$)}

The COG$_4$ algorithm shares some similarities with the two-strip COG. It has
a drop in the probability at $x\approx 0$ for sufficiently wide signal distributions.
Following our precedent convention, the ordering of the strips is $5,3,2,1,4$ with
the reference system centered in the middle of the strip $\# 2$. The strip $\# 4$
has a distance of 2 from the origin of the reference system, the strip $\# 5$ has
a distance -2. With these positions, the COG$_4$ algorithm is defined as:
\begin{equation}\label{eq:equation_12}
    x_{g_4}=\Big(\frac{2x_4+x_1-x_3}{x_1+x_2+x_3+x_4}\Big)\theta(x_4-x_5)+
    \Big(\frac{x_1-x_3-2 x_5}{x_1+x_2+x_3+x_5}\Big)\theta(x_5-x_4) \,.
\end{equation}
\noindent
Before the study of the complete PDF for COG$_4$, it is better to study an its partial
form, limiting to the first part of $x_{g_4}$ ($x$ in the following) neglecting
the $\theta$-function.

\subsection{A partial expression for COG$_4$ and its PDF for Gaussian noise}

Similarly to the case of COG$_3$, the simultaneous presence of the
$\theta$-function and the ratio of random variables does not
allow to find a closed form of the PDF for Gaussian noise and approximations
must be found. Instead, the partial form of COG$_4$, without the $\theta$-function,
has an analytical expression, useful to select a good
approximation for the complete PDF.
\begin{equation}\label{eq:equation_13}
   P_{xg_4}^r(x)= \int_{-\infty}^{+\infty}\,\mathrm{d} x_1\,
   \mathrm{d} x_2\, \mathrm{d} x_3\, \mathrm{d} x_4 P_1(x_1) P_2(x_2) P_3(x_3) P_4(x_4)
   \Big\{\delta\big(x-\frac{2x_4+ x_1-x_3}{x_1+x_2+x_3+x_4}\big)\Big\}\,.
\end{equation}
\noindent
The substitution of variables, $\xi=2 x_4+x_1-x_3$, $z=x_1+x_2+x_3+x_4$, $\beta=x_3$
and $\gamma=x_4$ with the Jacobian-determinant equal one, gives the integral of
the Dirac-$\delta$ function in the form:
\begin{equation}\label{eq:equation_14}
    P_{xg_4}^r(x)= \int_{-\infty}^{+\infty}\, P_1(z\,x+\beta-2\gamma)
    P_2(z(1-x)+\gamma-2\beta) P_3(\beta) P_4(\gamma)\, |z|\,\mathrm{d}z\,
    \mathrm{d} \beta\, \mathrm{d} \gamma\,.
\end{equation}
\noindent
This equation is general and defined for any signal PDF.
We calculate for a Gaussian additive noise supposed to perturb the signal collected
by the strips with the form of equation~\ref{eq:equation_5}.
The Gaussian integrals have now analytic expressions. The absolute value of $z$ obliges
to split the $z$-integral in two parts that introduces a function $\mathrm{Erf}$.
With the definition of $E_4^+$, $X_4^+$ and $\Sigma_4^+$ as auxiliary variables:
\begin{equation*}
   E_4^+=(a_1+a_2+a_3+a_4)\ \ \ \  X_4^+=\frac{(2a_4+a_1-a_3)}{E_4^+} \ \ \ \
   \Sigma_4^+= [(2-x)^2\sigma_4^2+(1-x)^2\sigma_1^2+x^2\sigma_2^2+(1+x)^2\sigma_3^2]\,.
\end{equation*}
The expressions of the results acquire a form with strong analogies with
the COG$_3$ PDFs. The main term has a Gaussian-like maximum centered on $X_4^+$
and the rest of the term contains a typical structure of differences. These
differences are the noiseless COG$_4$ calculated with the reference system
centered on the strip $\# 4$, on the strip $\# 3$, on the strip $\# 2$ and on
the strip $\# 1$ (a lucky suggestion of the MATHEMATICA function "FullSimplify").
The integration of equation~\ref{eq:equation_14} gives two terms, the first term is the
main term $P_{xg4}^{r\,a}(x)$:
\begin{equation}\label{eq:equation_15}
\begin{aligned}
    &P_{xg4}^{r\,a}(x)=\exp\Big[-(x-X_4^+)^2\frac{(E_4^+)^2}{2\Sigma_4^+}\Big]\\
     &\frac{(E_4^+)}{\sqrt{2\pi(\Sigma_4^+)^3\,\,}}{\Big[(2-X_4^+)(2-x)\sigma_4^2+
     (1-X_4^+)(1-x)\sigma_1^2+\,X_4^+\,x\sigma_2^2\,+
    \,(1+X_{4}^+)(1+x)\sigma_3^2\Big]}\\
    &\mathrm{Erf}\Big[\frac{(E_4^+)\big[(2-X_4^+)(\,2-x\,)\sigma_4^2\,+\,(1-X_4^+)\,(\,1-x\,)
    \,\sigma_1^2+\,X_4^+\,x\sigma_2^2\,+
    \,(1+X_4^+)\,(\,1+x\,)\,\sigma_3^2]}{\sqrt{2\,\Sigma_4^+\,
    [9\sigma_3^2\sigma_4^2+\sigma_1^2(\sigma_2^2+4\sigma_3^2+
    \sigma_4^2)+\sigma_2^2(\sigma_3^2+4\sigma_4^2)\big]}}\Big]\\
\end{aligned}
\end{equation}
\noindent
As previously, the second term $P_{xg4}^{r\,c}(x)$ is called Cauchy term, it survives when  all
the $a_i$ are zero:
\begin{equation}\label{eq:equation_16}
    \begin{aligned}
    P_{xg4}^{r\,c}(x)=
    \!\exp\!&\big[\!-\!\frac{(E_4^+)^2(2-X_4^+)^2\sigma_4^2\!+\!(1-X_4^+)^2\!\sigma_1^2+\!(X_4^+)^2\sigma_2^2\!+
    \!(1+X_4^+)^2\!\sigma_3^2}{{2\,
    [9\sigma_3^2\sigma_4^2+\sigma_1^2(\sigma_2^2+4\sigma_3^2+\sigma_4^2)+\sigma_2^2(\sigma_3^2+4\sigma_4^2)]}}\big]\\
    &\frac{1}{\pi\,\Sigma_4^+}{\sqrt{9\sigma_3^2\sigma_4^2+\sigma_1^2(\sigma_2^2+
    4\sigma_3^2+\sigma_4^2)+\sigma_2^2(\sigma_3^2+4\sigma_4^2)\,\,\,}}\,.
    \end{aligned}
\end{equation}
This form of the Cauchy term is different from the standard Cauchy PDF, but it
has similar properties. Its cumulative distribution
is an arctangent function as for a Cauchy PDF.

\noindent
The other part of the COG$_4$ PDF without the $\theta(x_5-x_4)$ is:
\begin{equation}\label{eq:equation_17}
     P_{xg_4}^l(x)= \int_{-\infty}^{+\infty}\,\mathrm{d} x_1\,
   \mathrm{d} x_2\, \mathrm{d} x_3\, \mathrm{d} x_4 P_1(x_1) P_2(x_2) P_3(x_3) P_5(x_5)
   \Big\{\delta\big(x-\frac{ x_1-x_3-2 x_5}{x_1+x_2+x_3+x_5}\big)\Big\}\,.
\end{equation}
The equation of $P_{xg_4}^l$ can be obtained from that of $P_{xg_4}^r$
with the substitutions $x_4\Leftrightarrow x_5$, $x_1 \Leftrightarrow x_3$
and $x\rightarrow -x$.
Thus, the insertion of them in equation\ref{eq:equation_15}, in
equation\ref{eq:equation_16} and in the definitions of the auxiliary variables
gives $P_{xg_4}^{l\,a}$ and $P_{xg_4}^{l\,c}$.

\subsection{The complete PDF for the COG$_4$ with the strip selection}

The first part to be studied is the term:
\begin{equation}\label{eq:equation_18}
   P_{xg_4}^R(x)= \int_{-\infty}^{+\infty} \mathrm{d} x_1\,\cdots
   \mathrm{d} x_5 P_1(x_1) P_2(x_2) P_3(x_3) P_4(x_4) P_5(x_5)
   \Big\{\delta\big(x-\frac{2x_4+ x_1-x_3}{x_1+x_2+x_3+x_4}\big)\theta(x_4-x_5)\Big\}\,.
\end{equation}
\noindent
The substitution of variables, $\xi=2 x_4+x_1-x_3$, $z=x_1+x_2+x_3+x_4$, $\beta=x_3$,
$\gamma=x_4$ and $\psi=x_5$ with their Jacobian-determinant equal to one, allows a direct
integration of the Dirac $\delta$-function and to account for the $\theta(x_4-x_5)$.
\begin{equation}\label{eq:equation_19}
    P_{xg_4}^R(x)= \int_{-\infty}^{+\infty}\,\mathrm{d}z\,\mathrm{d} \beta\,
    \mathrm{d} \gamma\, P_1(z\,x+\beta-2\gamma)
    P_2(z(1-x)+\gamma-2\beta) P_3(\beta) P_4(\gamma)\, |z|\int_{-\infty}^\gamma P_5(\psi)
    \, \mathrm{d} \psi.
\end{equation}
\noindent
With another substitution $\psi=\gamma+\delta$ the equation~\ref{eq:equation_19}
becomes:
\begin{equation}\label{eq:equation_20}
    P_{xg_4}^R(x)=\int_{-\infty}^0 \mathrm{d} \delta \int_{-\infty}^{+\infty}
    \,\mathrm{d}z\,\mathrm{d} \beta\, \mathrm{d} \gamma\, P_1(z\,x+\beta-2\gamma)
    P_2(z(1-x)+\gamma-2\beta) P_3(\beta) P_4(\gamma)P_5(\gamma+\delta)\, |z|
    \, .
\end{equation}
\noindent
The second part of this PDF (the left side) can be obtained in similar way
with the corresponding variables or obtained with  the
substitutions $x_4\Leftrightarrow x_5$, $x_1 \Leftrightarrow x_3$ and
$x\rightarrow -x$.
These equations are general for any probability $P_j$, we
calculate an explicit approximate expression for Gaussian PDFs.

\subsection{Approximate expression for Gaussian noise}

The integral on $P_5$, as usual, creates many problems and impedes an
analytical expression for the PDF. It conflicts with the $|z|$ that requires
the splitting of the integral on $z$. It was proved few times that
the effect of the full account of $|z|$ is to assure the PDF to
be a positive function. This condition can be released (with a small error)
and reinserted at the end of the integration. To avoid an unmanageable number
of terms in MATHEMATICA computation, it is better to start from the simpler
PDF for the COG$_3$ of ref.~\cite{landi10} observing that its integral
coincides with that of equation~\ref{eq:equation_20} apart from a
translation in $\gamma$ ($-2\gamma$ for P$_1$ and $\gamma$ for P$_2$).
The addition of these translations to the main term of this
closed form of COG$_3$, and the neglect of its Erf-function, allows
the computing of the remain two integrals without an explosion of
the length of the MATHEMATICA outputs.
As expected, the main term of the integrals
of equation~\ref{eq:equation_20} is very
similar to that of equation~\ref{eq:equation_15} the essential differences are
the absence of the $\mathrm{Erf}$-function, giving $\mathrm{A\,Erf(A)}$ to the factor of
the exponential, and the presence of a $(1-\mathrm{Erf(M)})/2$

\begin{equation}\label{eq:equation_21}
    \begin{aligned}
    &P_{xg4}^{R\,a}(x)=\exp\Big[-\frac{(E_4^+)^2(X_4^+-x)^2}
    {2\,\Sigma_4^+}\Big]\\
    &{\Big[(2-X_4^+)(2-x)\sigma_4^2+(1-X_4^+)(1-x)\sigma_1^2+\,X_4^+\,x\sigma_2^2\,+
    \,(1+X_{4}^+)(1+x)\sigma_3^2\Big]}\\
    &\frac{(E_4^+)}{\sqrt{2\pi(\Sigma_4^+)^3\,\,}}\frac{\mathrm{1-Erf(M)}}{2}\\
    &\  \\
    \end{aligned}
\end{equation}
\noindent
where $\mathrm{M}$ is defined as:
\begin{equation}\label{eq:equation_22}
    \begin{aligned}
    & \mathrm{M}=\Big[\frac{-(a_4-a_5)\Sigma_3
    +(2-x)(a_1-a_3+2a_5-x(a_1+a_2+a_3+a_5))\sigma_4^2}
    {\sqrt{2\Sigma_4^+
    \big(\Sigma_4^+\,\sigma_5^2+\Sigma_3\,\sigma_4^2\big)}}\Big]\\
    &\Sigma_3=[(1-x)^2\sigma_1^2+x^2\sigma_2^2+(1+x)^2\sigma_3^2]\,.\\
    \end{aligned}
\end{equation}
\noindent
The factor $(1-\mathrm{Erf(M)})/2$ is essential to produce
the drop of the PDF for $x\approx 0$ with a smooth
transition toward $P_{xg4}^{L}(x)$ that dominates the for $x<0$.
Another term (Cauchy-like) is given by:
\begin{equation}\label{eq:equation_23}
    \begin{aligned}
    &P_{xg4}^{R\,C}(x)=\\
    &\exp\Big[-\frac{\!(E_4^+)^2\big(X_4^+-\,x\big)^2\!\sigma_5^2
    \!+\!\big(\!a_1\!-\!a_3\!+\!2a_5\!-\!x\!(\!a_1\!+\!a_2\!+\!a_3\!+
    \!a_5\!)\big)^2\!\sigma_4^2+\!(\!a_5\!-\!a_4\!)^2\Sigma_3}
    {2(\Sigma_3\sigma_4^2+\Sigma_4^+\,\sigma_5^2)}\!\Big]\\
    &\frac{\big[(x-1)\sigma_1^2+2x\sigma_2^2+3(1+x)\sigma_3^2\big]
    \,\sigma_4^2}{2\,\pi\,\Sigma_4^+\,  \sqrt{\Sigma_4^+\,\sigma_5^2+
    \Sigma_3\,\sigma_4^2\,\,}}\,.\\
    \end{aligned}
\end{equation}
\noindent
We continue to indicate the term $P_{xg4}^{R\,C}(x)$ as a Cauchy term even if
now it is a complicated function of $x$. A different order of integration in
equation~\ref{eq:equation_20} produces the main term with unimportant
differences in the argument of the Erf-function and a different Cauchy
term, always heavily suppressed by the exponential term.
The left parts $P_{xg4}^{L}(x)$ and $P_{xg4}^{L\,C}(x)$ are obtained
from the previous equations with the indicated substitutions.

\begin{figure} [h!]
\begin{center}
\includegraphics[scale=0.8]{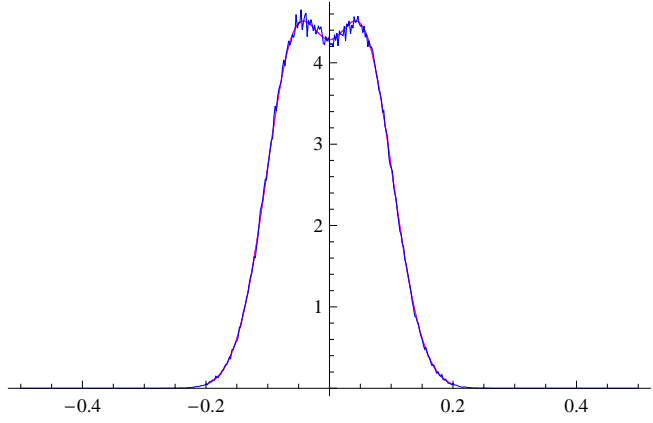}
\caption{\em  Plots of $P_{xg4}(x)$, the blue line is the histogram
of a simulation (orthogonal incidence in the strip center).
The red line is the numerical integration, and the magenta
line (overlapping the red line) is the result  of the previous
equations.
}\label{fig:figure_5}
\end{center}
\end{figure}

Figure~\ref{fig:figure_5} reports the comparison of the
approximate expressions  with a numerical integration
of equation and a simulation constructed with identical
data $\{a_i\}$ with the addition of a random Gaussian noise.

\section{The error PDF for five-strip COG}

The simplest form of the PDF for the five-strip COG (COG$_5$) is defined.
This COG$_5$ could have the border gaps, as usual for the COG algorithms with an odd
number of strips. In this case the gaps are relevant when the signal distribution
is appreciably larger than four strips. We neglect such case.
The general form of the COG$_5$ is:
\begin{equation}\label{eq:equation_24}
    x_{g_5}=\Big(\frac{2x_4+x_1-x_3-2 x_5}{x_1+x_2+x_3+x_4+x_5}\Big)\,.
\end{equation}
\noindent
As always, the COG$_5$ PDF is obtained with the integrals:
\begin{equation}\label{eq:equation_25}
   P_{xg_5}(x)= \int_{-\infty}^{+\infty} \mathrm{d} x_1\,\cdots
   \mathrm{d} x_5 P_1(x_1) P_2(x_2) P_3(x_3) P_4(x_4) P_5(x_5)
   \Big\{\delta\big(x-\frac{2x_4+ x_1-x_3-2x_5}{x_1+x_2+x_3+x_4+x_5}\big)\Big\}\,.
\end{equation}
\noindent
With the substitution of variables, $\xi=2 x_4+x_1-x_3-2 x_5$,
$z=x_1+x_2+x_3+x_4+x_5$, $\beta=x_3$,
$\gamma=x_4$ and $\psi=x_5$ and their Jacobian-determinant equal to one, the
integration in $\xi$ of the Dirac $\delta$-function gives:
\begin{equation}\label{eq:equation_26}
    P_{xg_5}(x)= \int_{-\infty}^{+\infty}\, P_1(z\,x+\beta-2\gamma+2 \psi)
    P_2(z(1-x)-2\beta+\gamma-3\psi) P_3(\beta) P_4(\gamma)\, P_5(\psi)\,  |z|
    \,\mathrm{d}z\,\mathrm{d} \beta\, \mathrm{d} \gamma\, \mathrm{d} \psi.
\end{equation}
\noindent
This PDF can be used with any probability $P_j(x_j)$, we specialize it for Gaussian
probabilities.

\subsection{The $P_{xg_5}(x) $ for the Gaussian noise}

The set of integrations in equation~\ref{eq:equation_26} have a
closed form with an evident correlation with the expressions of
the previously calculated PDFs. Due to the expressions we gave to the
other exact PDFs, it is easy to anticipate the analytical expression
of this PDF. A factor only ($F_5$ in the following) has a non-easy
relation to the precedent expressions.
This factor is contained in the Erf function and in the
Cauchy term. Even if the approximate
expressions without the Erf functions work very well and the Cauchy
term is negligible, for consistency this term must be calculated.
The Gaussian integrals of equation~\ref{eq:equation_26}
require further attentions to avoid the explosion of the number
of terms in the exponents. MATHEMATICA is able to handle and to
quickly simplify very long expressions as those of COG$_3$.
But, the lengths of the intermediate integrations become
excessive for $P_{xg_5}(x) $, to handle this case,  many definitions
of auxiliary variables must be introduced and few partial results
must be simplified.
The structure of equation ~\ref{eq:equation_26} is very
similar to equation~\ref{eq:equation_14} with the addition of
another integration on $\psi$. This set of integrals can be
calculated in succession with the last integral on $z$.
The factor $|z|$ introduces the Erf-function that
renders difficult further integrations.

To shorten the analytical expressions, the following auxiliary variables are very
helpful:
\begin{equation*}
   E_5=(a_1+a_2+a_3+a_4+a_5)\ \ \ \  X_5=\frac{(2a_4+a_1-a_3-2 a_5)}{E_5} \ \ \ \
\end{equation*}
and
\begin{equation*}
\begin{aligned}
    &\Sigma_5= [(2-x)^2\sigma_4^2+(1-x)^2\sigma_1^2+x^2\sigma_2^2+
    (1+x)^2\sigma_3^2+(2+x)^2\sigma_5^2]\ \ \ \ \ \ \ \ \ \ \ \ \ \ \ \ \ \\
    &F_5=[16\sigma_4^2\sigma_5^2+\sigma_3^2(9\sigma_4^2+\sigma_5^2)+\sigma_1^2(\sigma_2^2+4\sigma_3^2
    +\sigma_4^2+9\sigma_5^2)+\sigma_2^2(\sigma_3^2+4\sigma_4^2+4\sigma_5^2)]\\
\end{aligned}
\end{equation*}
the equation~\ref{eq:equation_26} for $P_{xg_5}(x) $ gives the most important
term with the expression:
%
%
%

\begin{equation}\label{eq:equation_27}
    P_{xg5}(x)=\exp\Big[-(x-X_5)^2\frac{E_5^2}{2\Sigma_5}\Big]\,\,
    \frac{E_5\,H_5}{\sqrt{2\pi\,\Sigma_5^3\,\,}}
    \mathrm{Erf}\Big[\frac{\textstyle{E_5\,H_5}}{\sqrt{2\,\Sigma_5\,
    F_5}}\Big]
\end{equation}

\[H_5=\textstyle{{\Big[(2-X_5)(2-x)\sigma_4^2+(1-X_5)(1-x)\sigma_1^2+\,X_5\,x\sigma_2^2\,+
    \,(1+X_5)(1+x)\sigma_3^2+\,(2+X_5)(2+x)\sigma_5^2\Big]}} \]
\noindent
The approximate expression neglecting the Erf-term can be written easily. The approximation
$A\,\mathrm{Erf}(A)\approx |A|$ works very well.
The Cauchy term is now:
\begin{equation}\label{eq:equation_28}
    P_{xg5}^{c}(x)=\frac{\sqrt{F_5}}{\pi\,\Sigma_5}
    \!\exp\!\Big[\!-\!\frac{(E_5)^2[(2-X_5)^2\sigma_4^2\!+\!(1-X_5)^2\!\sigma_1^2+\!(X_5)^2\sigma_2^2+
    (1+X_5)^2\!\sigma_3^2+(2+X_5)^2\sigma_5^2]}{{2\,
    F_5}}\Big]
    \,.
\end{equation}
This term has the functional dependence on $x$  in the denominator of
equation~\ref{eq:equation_28} as a polynomial of second degree. Its
cumulative function is always an arctangent as for the Cauchy PDF.

\section{Conclusions}

General expressions are reported for probability
density functions of the center of gravity algorithms
with three, four and five discretized values (strips for
silicon detectors).
These expressions are calculated with the method illustrated
in a previous publication. Whenever possible, exact
expressions for Gaussian noise are calculated. Approximate forms are
constructed  when the analytical integrations are impossible.
The quality of these approximations are evaluated
with comparisons with numerical integrations and data simulations.
For the complete three strips and four strips probability distributions,
exact analytical integrals are obtained neglecting the selection
of the leading strips. These exact, but incomplete, expressions
are useful to select good approximations. Even if our attention
is directed to the center of gravity as a positioning algorithm
in silicon micro-strip trackers, evidently these probability
distributions have wider generality.


\end{document}